\documentclass[10pt,journal,cspaper,compsoc]{IEEEtran}

\RequirePackage[colorlinks]{hyperref}
\hypersetup{citecolor=blue, linkcolor=blue}

\usepackage{overpic}
\usepackage{algorithm,algorithmic}
\usepackage{amsmath,amssymb,amsfonts,mathrsfs,graphicx,enumerate,amsthm}

\newcommand\diff{\,{\mathrm d}}
\newcommand*\Laplace{\mathop{}\!\mathbin\bigtriangleup}

\begin{document}

\title{Conformal Surface Morphing with Applications on Facial Expressions}

\author{Mei-Heng~Yueh, Xianfeng~David~Gu, Wen-Wei~Lin, Chin-Tien~Wu, and~Shing-Tung~Yau}

\IEEEcompsoctitleabstractindextext{%
\begin{abstract}
Morphing is the process of changing one figure into another. Some numerical methods of 3D surface morphing by deformable modeling and conformal mapping are shown in this study. It is well known that there exists a unique Riemann conformal mapping from a simply connected surface into a unit disk by the Riemann mapping theorem. The dilation and relative orientations of the 3D surfaces can be linked through the M\"obius transformation due to the conformal characteristic of the Riemann mapping. On the other hand, a 3D surface deformable model can be built via various approaches such as mutual parameterization from direct interpolation or surface matching using landmarks. In this paper, we take the advantage of the unique representation of 3D surfaces by the mean curvatures and the conformal factors associated with the Riemann mapping. By registering the landmarks on the conformal parametric domains, the correspondence of  the mean curvatures and the conformal factors for each surfaces can be obtained. As a result, we can construct the 3D deformation field from the surface reconstruction algorithm proposed by Gu and Yau. Furthermore, by composition of the M\"obius transformation and the 3D deformation field, the morphing sequence can be generated from the mean curvatures and the conformal factors on a unified mesh structure by using the cubic spline homotopy. Several numerical experiments of the face morphing are presented to demonstrate the robustness of our approach.
\end{abstract}

}

\maketitle

% is not selected <OR> if conference mode is selected - because compsoc
% conference papers position the abstract like regular (non-compsoc)
\IEEEdisplaynotcompsoctitleabstractindextext

% \fi
\IEEEpeerreviewmaketitle

\section{Introduction}

\IEEEPARstart{T}{he} metamorphosis between two objects is commonly called "morphing". It is the process of changing one figure into another. In recent years, image morphing techniques have been widely used in the entertainment industry. Many techniques have been developed to achieve a desired morphing effect. \cite{W1998}

In 2D image morphing, S.-Y. Lee et al. \cite{1994LCHS} proposed a technique that generates a $C^1$-continuous and one-to-one deformation from the two-dimensional positional constraints by using a deformable surface model. The transition behavior can be controlled by assigning the transition curves for selected points on an image. 
Similarly, A. Gregory et al. \cite{1999GSLML} proposed a method of geometry morphing, which creates a morph by defining morphing trajectories between the feature pairs and by interpolating them across the merged polyhedron.
Attributed to these technologies, the user can achieve a satisfactory visual effect by simply setting the correspondence between graphical features. 
In particular, for images of human faces, it is possible to pick the feature points of the face automatically. 
V. Zanella et al. \cite{ZRVR2009} use the so-called "Active Shape Models" \cite{CTCG1995} to find the facial features in the 2D images and perform the morphing of face images in frontal view automatically. 
To make the metamorphosis look realer, the morphing path must be constrained by some physical laws. Researchers, such as Verbeek et al. \cite{1993VV}, Y. Bao et al. \cite{2005BGQ} and L. Younes et al. \cite{2001CY}, proposed methods of determining the morphing paths by optimizing an energy functional, which characterizes the intrinsic deformation of the surface away from its rest shape. 
Althrough 2D image morphing technique has pretty mature, 3D image morphing remains challenges, especially when the virtual real morphing effects are desired.
In addition, in order to achieve a satisfactory visual effect, the texture images also need to be computed in the process of visualization. 

On the other hand, with the advance of the three-dimensional imaging technology, surface morphing in 3D has become very important. 
Comparing to the 2D image matching problem, surface matching problem is much more difficult, since the surface matching involves the correspondence in $\mathbb{R}^3$ coordinates and the geometric information of images in $\mathbb{R}^3$ is far richer than images in 2D. 
D. DeCarlo et al. \cite{1996DG} proposed a method of generating smooth-looking transformations between pairs of surfaces that may differ in topology by specifying a sparse control mesh on each surface and by associating each face in one control mesh with a corresponding face in the other.
Y.-S. Liu et al. \cite{2011LYM} proposed a morphing method that works well even when two input 3D triangle meshes have very different shapes by creating consistent meshes for the original source and target models.
Schr\"oder et al. \cite{2005LDRS} proposed a variational approach based on minimizing bending and stretching in which corresponding feature points and line segments are matched.
E. Jeong et al. \cite{JYLALG2003} proposed a feature-based morphing technique for two objects equipped with surface light fields by using spherical embedding of meshes. 
M. Yang et al. \cite{YWZ2013} proposes a realistic 3D morphing method based on GPU for real-time animation of facial expressions by rendering the real texture on the digital character. 

In this study, we propose an efficient way to obtain the desired morphing effects. Our goal is to generate the morphing sequence between human faces in which features of the given surfaces are maintained during the morphing process.
Moreover, in order to control digital models and perform digital acting via real agents, we also construct the single mesh of human faces through the correspondence matching.
In our approach, we utilize the conformal mapping technique. A M\"obius transformation and a deformation from a modified thin-plate model are computed to achieve high accurate feature correspondence.
A single mesh based on geodesic paths among feature points is computed very efficiently. With the help provided by this single mesh, the texture information and the geometric information, such as the conformal factor $\lambda_i$ and the mean curvature $H_i, i=1 \cdots n$, of each frame in the morphing sequence can be easily interpolated from the given surfaces through various homotopy methods. Then the 3D images of each intermediate frame can be computed from the reconstruction algorithm proposed by Gu and Yau. 
Our method guarantees the piecewise smoothness of homotopy path. So, we can avoid the vibration of each frames caused by the unstableness of the 3D camera. In addition, we merely require 10\% the number of original frames to simulate the real motion of human facial expressions, so that the size of the information data is significantly reduced but the resolution of the 3D images does not decrease. 

In the following, we review the computation of the Riemann conformal mapping in section \ref{Sec:ICRM}. We present our surface matching technique in section \ref{Sec:SMMGE} and introduce the way of homotopy in section \ref{Sec:SMH}. After that, we show how we reconstruct the morphing sequence in section \ref{Sec:SR} and demonstrate some morphing results in section \ref{Sec:SM3D}. 

\section{Idea for Computing Conformal Mapping} \label{Sec:ICRM}
The Riemann conformal mapping plays an important role in the surface matching. In this section, we briefly review the method of computing Riemann conformal mapping.

\subsection{Spherical Conformal Mapping}
\begin{figure}
\centering
\includegraphics[height=2.5cm]{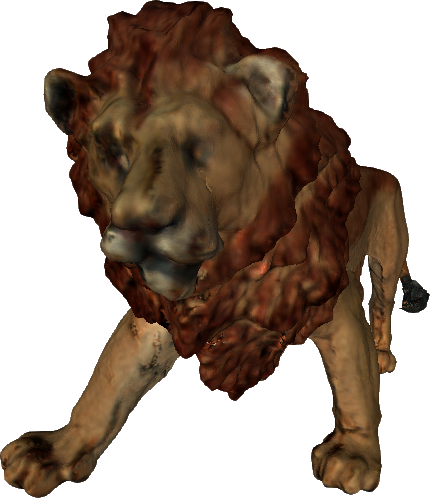}
\hspace{1cm}
\begin{overpic}[height=2.5cm]{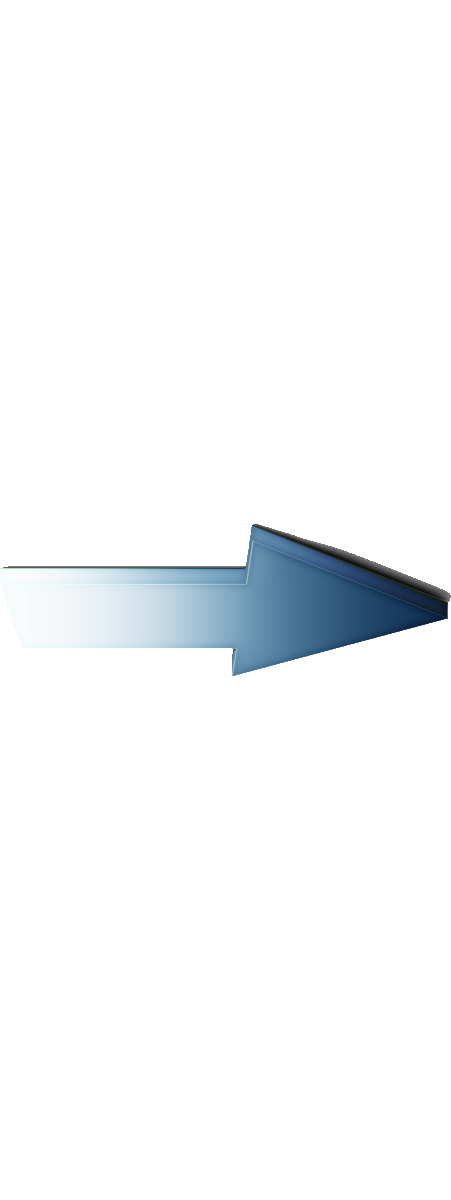}
\put(-50,60){Spherical Conformal}
\end{overpic}
\hspace{1cm}
\includegraphics[height=2.5cm]{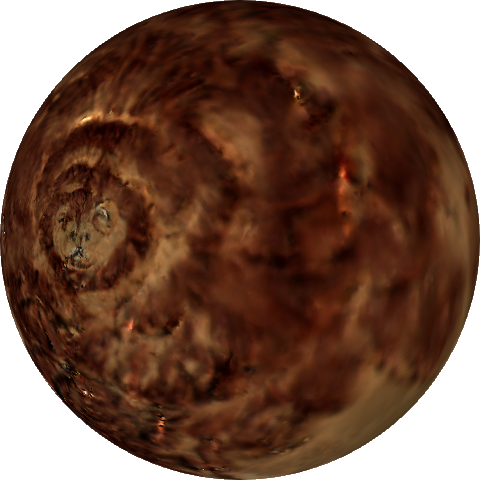}
\caption{Spherical conformal mapping}
\end{figure}

The spherical conformal mapping is first computed by Gu and Yau \cite{2004GWCTY} in 2003. 
The idea is based on minimizing the harmonic energy through a nonlinear heat diffusion process as following: 
Suppose the desired map is $\varphi : \mathcal{M}\rightarrow \mathbb{S}^2.$ Let $\varphi(v)$ and $\mathbf{n}(\varphi(v))$ denote the image of the vertex $v\in \mathcal{M}$ and the normal at $\varphi(v),$ respectively. The normal and tangent components of $\Laplace \varphi$ are defined as
\begin{align*}
(\Laplace \varphi(v))^\perp = \langle \Laplace \varphi(v), \mathbf{n}(\varphi(v)) \rangle \mathbf{n}(\varphi(v))
\end{align*}
and
\begin{align*}
(\Laplace \varphi(v))^\parallel = \Laplace \varphi(v) - (\Laplace \varphi(v))^\perp,
\end{align*}
respectively. The harmonic map is then computed by minimizing the harmonic energy associated with $\varphi$ through the nonlinear heat diffusion process 
\begin{align*}
\frac{\diff \varphi}{\diff t} &= -(\Laplace \varphi)^\parallel,
\end{align*}
with the constrain $\varphi(\mathcal{M},t)\in \mathbb{S}^2$.
The equation can be written in the following form,
\begin{eqnarray*}
\frac{\diff \varphi}{\diff t} &=& -(\Laplace \varphi)^\parallel \\
&=& -\left( \Laplace \varphi - \left\langle \Laplace \varphi, \mathbf{n}(\varphi) \right\rangle \mathbf{n}(\varphi) \right)\\
&=& -\left( \Laplace \varphi - \left\langle \Laplace \varphi, \varphi \right\rangle \varphi \right),
\end{eqnarray*}
and can be solved numerically by using the quasi-implicit Euler method (QIEM) \cite{2013GHHLLY}, 
$$\left[ I + \delta t^{(m)} \left( K-D^{(m)} \right) \right] \varphi^{(m+1)} = \varphi^{(m)},$$
where, $K$ is the discrete Laplacian, $D^{(m)}$ is a diagonal matrix with
$$\left( D^{(m)} \right)_{ii} = \left\langle \left(K\,\varphi^{(m)} \left(v_i\right)\right), \varphi^{(m)}\left(v_i\right) \right\rangle.$$
The initial map $\varphi^{(0)}$ for the above iterative formula can be obtained from the Gauss map. 
It is well known that the efficiency of the explicit scheme is usually not satisfactory since the time step is generally very small due to the diffusive nature. The quasi-implicit Euler method \cite{2013GHHLLY} proposed by W.-W. Lin et al. is shown to be very robust on solving the nonlinear diffusion equation. 

\subsection{Riemann Conformal Mapping}
\begin{figure}
\centering
\begin{overpic}[height=2cm]{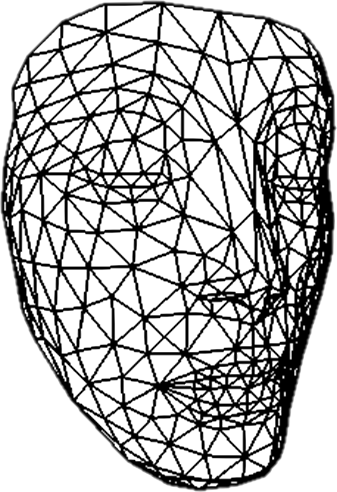}
\end{overpic}
\hspace{1.2cm}
\begin{overpic}[height=2cm]{images/arrow_right.png}
\put(-30,60){Riemann Map}
\end{overpic}
\hspace{1.2cm}
\begin{overpic}[height=2cm]{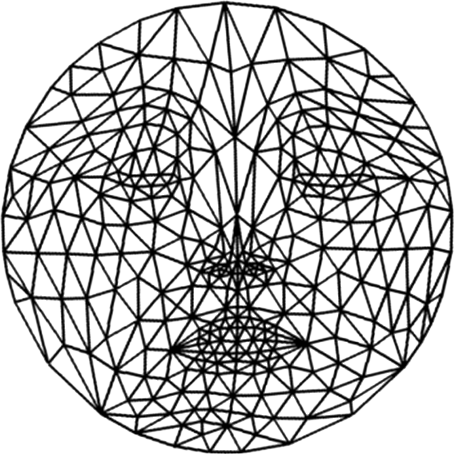}
\end{overpic} \\
\vspace{0.7cm}
\begin{overpic}[height=0.8cm]{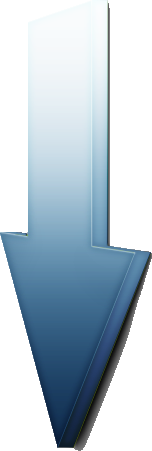}
\put(-130,40){Double Covering}
\end{overpic}
\hspace{4.5cm}
\begin{overpic}[height=0.8cm]{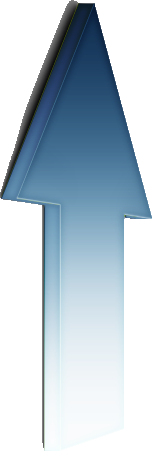}
\put(-230,20){Stereographic Projection}
\put(-160,-40){M\"obius Transform}
\end{overpic}
\\
\vspace{0.7cm}
\begin{overpic}[height=2cm]{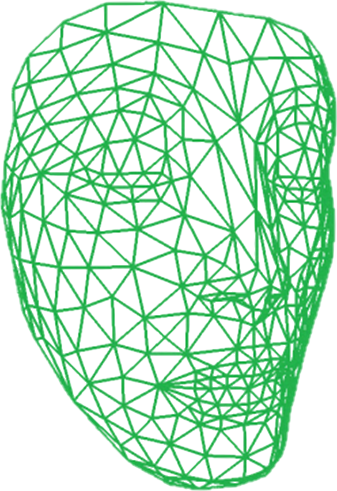}
\put(2,-2){
\includegraphics[height=2cm]{images/face_mesh.png}
}
\end{overpic}
\hspace{1.3cm}
\begin{overpic}[height=2cm]{images/arrow_right.png}
\put(-55,69){Spherical Conformal}
\end{overpic}
\hspace{1.3cm}
\begin{overpic}[height=2cm]{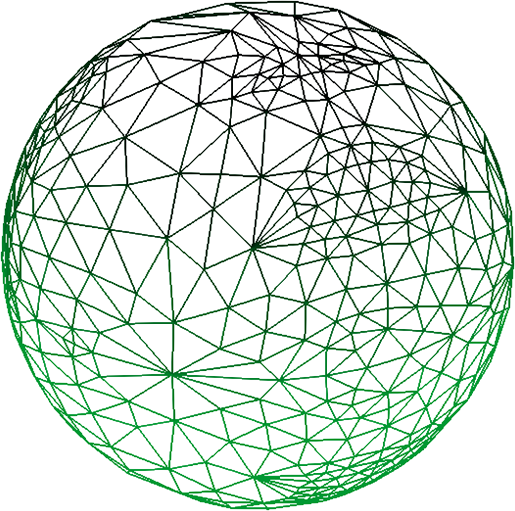}
\end{overpic}
\caption{Idea for computing Riemann conformal mapping}
\label{fig:RiemannMappingIdea}
\end{figure}

The idea for computing Riemann conformal mapping, shown in Figure \ref{fig:RiemannMappingIdea}, was first proposed by Gu and Yau \cite{2008GY}. For a given surface $\mathcal{M}$ with single boundary, we make it into a closed surface $\overline{\mathcal{M}}$ by using the double covering technique. Then we compute the spherical conformal mapping of $\overline{\mathcal{M}}$. Finally, we cut out the semi-sphere along the equator and map the unit semi-sphere onto the unit disk conformally by using the stereographic projection. 

\subsection{Numerical Results}
\begin{figure}
\centering
\begin{tabular}{cccc}
\includegraphics[height=2cm]{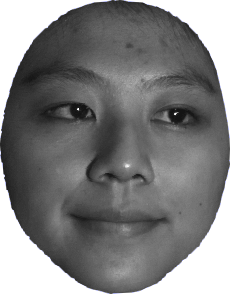} &
\includegraphics[height=2cm]{images/arrow_right.png} &
\includegraphics[height=2cm]{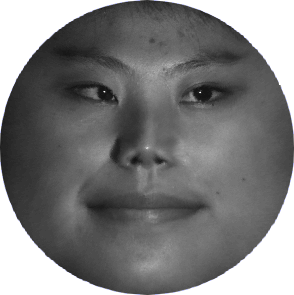} &
\includegraphics[height=2cm]{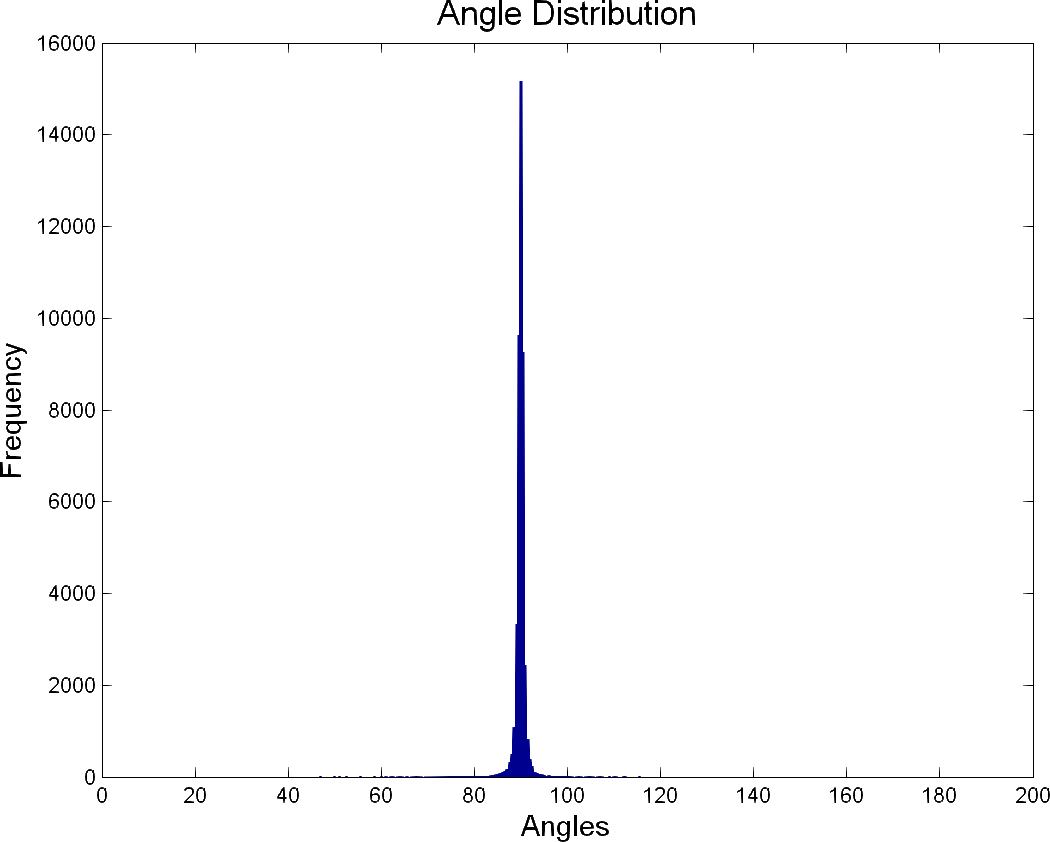} \\
\includegraphics[height=2cm]{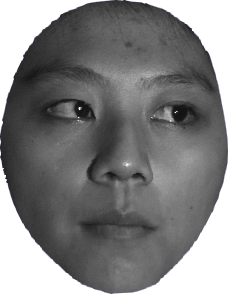} &
\includegraphics[height=2cm]{images/arrow_right.png} &
\includegraphics[height=2cm]{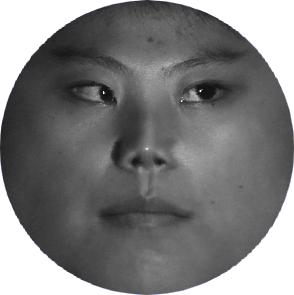} &
\includegraphics[height=2cm]{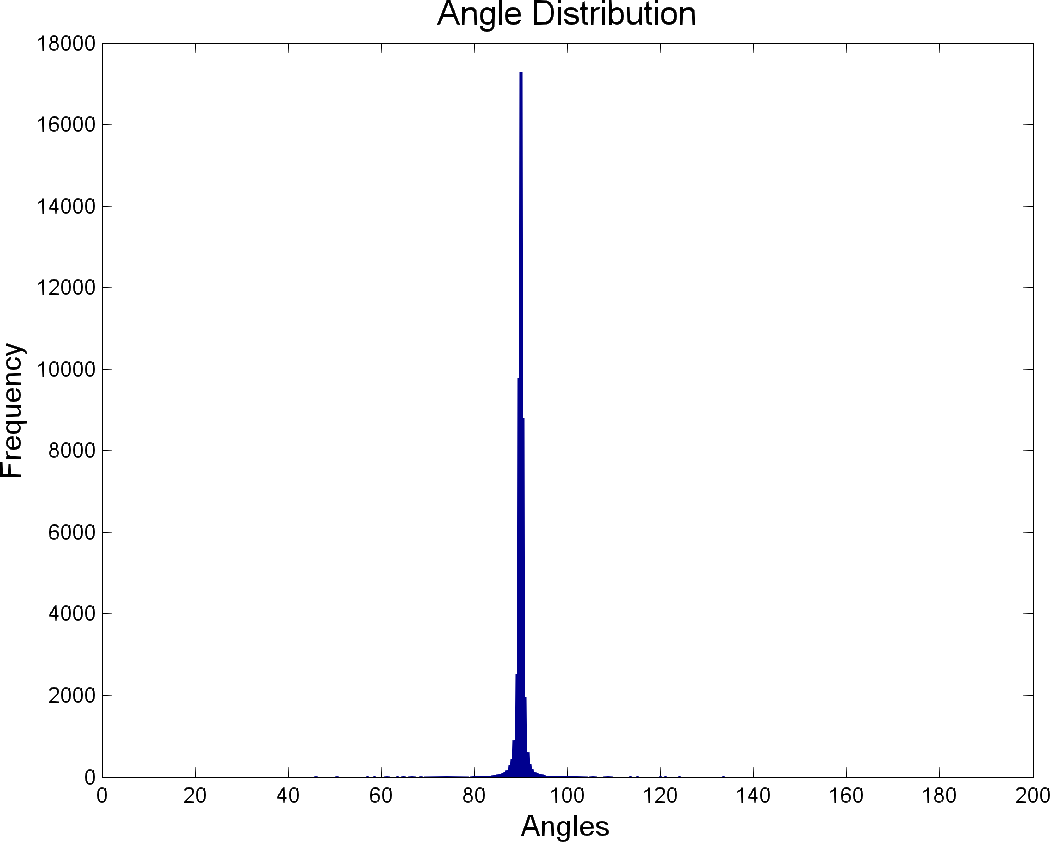} 
\end{tabular}
\caption{The results of the Riemann conformal mapping}
\label{fig:ConformalResults}
\end{figure}

We demonstrate the robustness of the quasi-implicit Euler method by computing the Riemann conformal mappings of human facial expressions. Two different facial expressions and the associated conformal mappings are shown in Figure \ref{fig:ConformalResults}. 
In order to check the conformality of the QIEM, we paste the checkerboard grid on the image of the Riemann conformal mapping $\varphi(\mathcal{M})$, and put it back to the surface $\mathcal{M}$ by using the inverse of the Riemann conformal mapping $\varphi^{-1}$. If the mapping is angle-preserving, every angle should be nearly 90 degrees. 
The histograms of the angle distribution, shown in Figure \ref{fig:ConformalResults}, indicate that the QIEM is very accurate on angle-preserving. 

\section{Surface Matching with Minimal Global Error} \label{Sec:SMMGE}
Surface matching plays a critical role in surface morphing.
L. Younes et al. \cite{2001CY} proposed a simple method to interpolate 2D landmark matching by constructing a diffeomorphism between two domains.
When it comes to $\mathbb{R}^3$ surface matching, it would be much more difficult. However, with the Riemann conformal mappings, we reduce the $\mathbb{R}^3$ surface matching problem into the unit disk matching problem. Hence we can apply the similar idea of the 2D landmark matching to the $\mathbb{R}^3$ surface matching.

For the $\mathbb{R}^3$ surface matching, X. Gu et al. \cite{2004GWCTY} use landmarks to obtain the optimal M\"obius transformation as the matching function between the spherical conformal mappings of two brains. When it comes to the matching between two different facial expressions, a slight twist is allowed since the transformation of facial expressions is actually not conformal. In the following, we propose to match the landmarks of each facial expressions by composition of the M\"obius transformation and deformation from the plate matching.

\begin{figure}
\centering
\begin{overpic}[height=2cm]{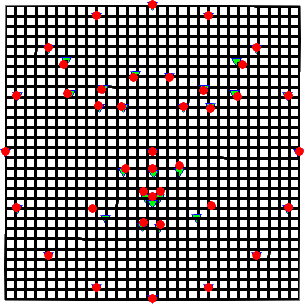}
\put(-160, 0){
\includegraphics[height=2cm]{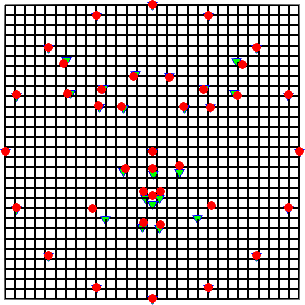}
}
\put(-50, 0){
\includegraphics[height=2cm]{images/arrow_right.png}
}
\put(-40, 60){$f_m$}
\put(110, 0){
\includegraphics[height=2cm]{images/arrow_right.png}
}
\put(120, 60){$f_g$}
\put(150, 0){
\includegraphics[height=2cm]{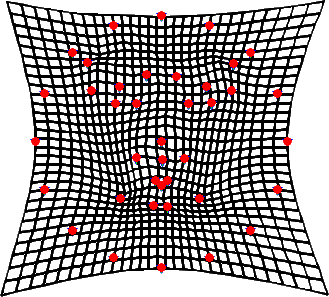}
}
\end{overpic}
\caption{The deformation field of $f_\mathbb{D}=f_m\circ f_g$ where $f_m$ is the optimal M\"obius transformation (OMT) and $f_g$ is the global matching function (GMF)}
\label{fig:MatchingResult}
\end{figure}

We assume that the correspondence of the boundary points of two different human faces $S_a$ and $S_b$ are known. The assumption can be realized by putting markers on the boundary of the human faces during scanning.
Suppose $m$ landmarks are selected for each facial expression and denoted by
\begin{align*}
\left\{ p^{(i)} \equiv \left(p^{(i)}_1, p^{(i)}_2, p^{(i)}_3\right) \right\}_{i=1}^m\subset S_a
\end{align*}
and
\begin{align*}
\left\{ q^{(i)} \equiv \left( q^{(i)}_1, q^{(i)}_2, q^{(i)}_3 \right) \right\}_{i=1}^m\subset S_b.
\end{align*}
Our purpose is to construct a matching function $f:S_a\to S_b,$ where $f\left(p^{(i)}\right)\approx q^{(i)},$ $i=1,\ldots,m.$
First, each surface is mapped to the unit disk $\mathbb{D}$ by the Riemann conformal mappings
$$\varphi_a:S_a\to\mathbb{D} \hspace{0.5cm}\textrm{and}\hspace{0.5cm} \varphi_b:S_b\to\mathbb{D}.$$
We denote $\varphi_a\left(p^{(i)}\right)$ by $p_{\mathbb{D}}^{(i)} \equiv \left( p_{\mathbb{D},1}^{(i)}, p_{\mathbb{D},2}^{(i)}\right)$ and 
$\varphi_b\left(q^{(i)}\right)$ by $q_{\mathbb{D}}^{(i)} \equiv \left( q_{\mathbb{D},1}^{(i)}, q_{\mathbb{D},2}^{(i)}\right)$, respectively, $i=1,\ldots,m.$
\begin{figure}
\centering
\begin{overpic}[height=2.8cm]{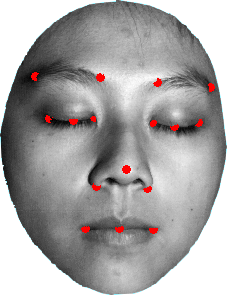}
\put(0,100){$S_a$}
\end{overpic}
\hspace{0.7cm}
\begin{overpic}[height=3cm]{images/arrow_right.png}
\put(10,60){$f$}
\end{overpic}
\hspace{0.7cm}
\begin{overpic}[height=2.8cm]{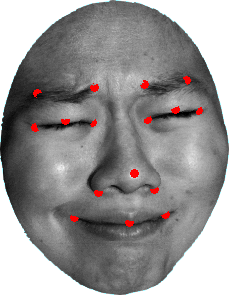}
\put(0,100){$S_b$}
\end{overpic} \\
\vspace{0.3cm}
\begin{overpic}[height=1cm]{images/arrow_down.png}
\put(-70,60){$\varphi_a$}
\end{overpic}
\hspace{4.5cm}
\begin{overpic}[height=1cm]{images/arrow_down.png}
\put(-70,60){$\varphi_b$}
\end{overpic}
\\
\begin{overpic}[height=2.5cm, scale=0.25, tics=10, angle=0]{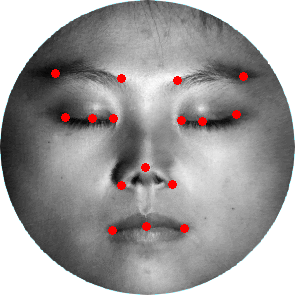}
\end{overpic}
\hspace{0.5cm}
\begin{overpic}[height=3cm]{images/arrow_right.png}
\put(10,60){$f_{\mathbb{D}}$}
\end{overpic}
\hspace{0.5cm}
\begin{overpic}[height=2.5cm, scale=0.25, tics=10, angle=0]{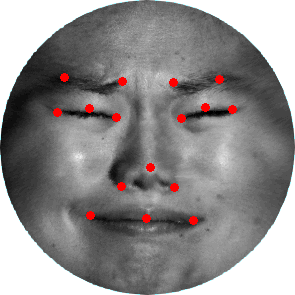}
\end{overpic}
\caption{The idea of surface matching}
\label{fig:MatchingIdea}
\end{figure}
Next, a uniform grid points $c^{(j)}$ of an $n \times n$ checkerboard $\mathcal{C}$ ranged on $\mathbb{A}=\left[-1,1\right]\times\left[-1,1\right]$, $j = 1, 2, \ldots, n^2$, is lay on $\mathbb{D}$. 
In the thin-plate model, the deformation field is approximated by the span of the Green functions $r^2\log r$ of the bending operator at each grid point where $r$ is the distance between $c^{(j)}$ and $x\in\mathbb{A}$.
Therefore the matching function between the unit disks $f_{\mathbb{D}}:\mathbb{D}\to\mathbb{D}$ is defined by $f_{\mathbb{D}}\left( x_1, x_2 \right) = \left( f^{(1)}_{\mathbb{D}}\left( x_1, x_2 \right), f^{(2)}_{\mathbb{D}}\left( x_1, x_2 \right) \right)$ with $x=\left( x_1,x_2 \right)$ and
\begin{align*}
f^{(k)}_{\mathbb{D}}\left( x_1, x_2 \right) = \sum_{j=1}^{n^2} \left( \alpha^{(k)}_j \left\|x - c^{(j)}\right\|^2 \log \left\| x - c^{(j)} \right\| \right), 
\end{align*}
for $k = 1,2,$ where $\alpha^{(k)}_j$
are unknown coefficients, $j=1,\ldots,n^2$.
To determine these coefficients for matching landmarks on conformal parametric domain, we solve the least square problems
$$\arg\min_{\alpha^{(k)}} \left\| 
S\alpha^{(k)} - q_{\mathbb{D},k}\right\|_2, \,\,k=1,2,$$
where
\begin{align*}
S_{ij} = &\frac{\lambda_t^{(j)}}{\lambda_o^{(j)}}
\left\| p_{\mathbb{D}}^{(i)} - c^{(j)} \right\|^2 \log\left\| p_{\mathbb{D}}^{(i)} - c^{(j)} \right\|,\\
 &i = 1,\ldots, m,\, j = 1,\ldots, n^2,
\end{align*}
$\lambda_a^{(j)}$ and $\lambda_b^{(j)}$ are the conformal factors, resulted from the Riemann conformal mappings $\varphi_a$ and $\varphi_b$, at $c^{(j)}$, respectively, and
\begin{align*}
\alpha^{(k)} = \left[ \alpha_1^{(k)}, \ldots, \alpha_{n^2}^{(k)} \right]^\top,
\hspace{0.5cm} 
q_{\mathbb{D},k} = \left[ q_{\mathbb{D},k}^{(1)}, \ldots, q_{\mathbb{D},k}^{(m)} \right]^\top.
\end{align*}
The least square problem can be easily solved by QR method when $n^2\leq m$ or solved by Tikhonov regularization \cite{1996EHN} 
\[
\left( \varepsilon I + S^\top S  \right) \alpha^{(k)} = S^\top q_{\mathbb{D},k}
\]
when $n^2>m.$
Then the matching function between $S_a$ and $S_b$ can be obtained by taking $f=\varphi_b^{-1}\circ f_\mathbb{D}\circ\varphi_a.$ 
Figure \ref{fig:MatchingResult} shows the result of global matching and Table \ref{Tab:Comparison} shows the comparison between the optimal M\"obius transformation (OMT) and the optimal M\"obius transformation with the global matching function (OMGMF). 

There is a variety ways to measure the distortion of the matching function, such as Euclidean norm on the $\mathbb{R}^3$ space, the infinite norm on the unit disk, etc. 
In order to verify that the matching pairs $p_\mathbb{D}^{(i)}$ are mapped to $q_\mathbb{D}^{(i)}$ by $f_{\mathbb{D}},$ $i=1, \ldots, m,$ we define the conformal matching energy
\begin{align*}
E_{\mathbb{D}}(f_\mathbb{D}) = \sum_{i=1}^m \frac{\lambda_b^{(i)}}{\lambda_a^{(i)}}\left\| f_{\mathbb{D}}\left(p_\mathbb{D}^{(i)}\right) - q_\mathbb{D}^{(i)} \right\|_2^2, \,\, p_\mathbb{D}^{(i)}, q_\mathbb{D}^{(i)} \in \mathbb{D},
\end{align*}
which measures the square of Euclidean distance at each matching pair on the conformal disk. 
On another point of view, we define the local matching energy
\begin{align*}
E_{\rm{loc}}(f) = \sum_{i=1}^m \left\| f\left(p^{(i)}\right) - q^{(i)} \right\|_2^2, \,\,\, p^{(i)}\in S_a,\, q^{(i)} \in S_b,
\end{align*}
which directly measures the square of Euclidean distance at each matching pair in $\mathbb{R}^3$. 
In order to ensure that the global error is relatively small, we also compare the global matching energy, which is defined by 
\[
E(f) = \int_\mathbb{D} \left| f\left( S_a(x)\right) - S_b(x) \right|^2 \diff x. 
\] 
According to Table \ref{Tab:Comparison}, the global matching function significantly reduces the matching energies, which indicates that the global matching function works well on the surface matching. 
\begin{table}
\centering
\begin{tabular}{c|c|c}
\hline
     & OMT & OMGMF  \\
\hline\hline
$E_{\mathbb{D}}(f_{\mathbb{D}})$ & $3.8820\times 10^{-4}$ & $6.8836\times 10^{-5}$  \\
\hline
$E_{\rm{loc}}(f)$ & $1.5029 \times 10^{-2}$ &  $4.0888\times 10^{-3}$ \\
\hline
$E(f)$ & $4.8951 \times 10^{-5}$ &  $1.9131\times 10^{-5}$ \\
\hline
\end{tabular}
\caption{Comparison between the optimal M\"obius transformation (OMT) and the optimal M\"obius transformation with the global matching function (OMGMF)}
\label{Tab:Comparison}
\end{table}

%%%%%%%%%%%%%%%%%%%%%%%%%%%%%%%%%%%%%%%%%%%
%%%%%%%%%%%%%%%%%%%%%%%%%%%%%%%%%%%%%%%%%%%

\section{Surface Morphing} \label{Sec:SMH}
\begin{figure}
\centering
\begin{tabular}{cccc}
\includegraphics[height=2cm]{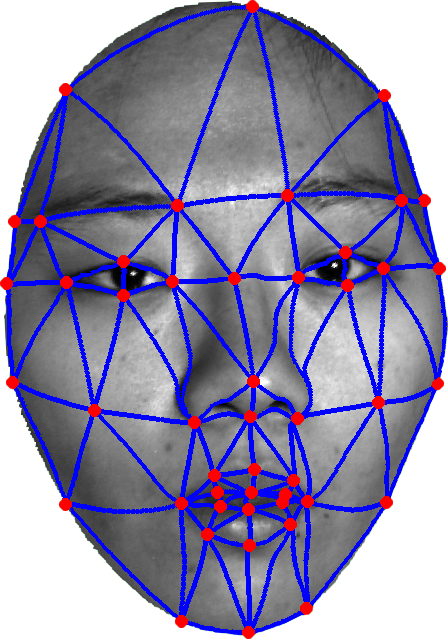} &
\includegraphics[height=2cm]{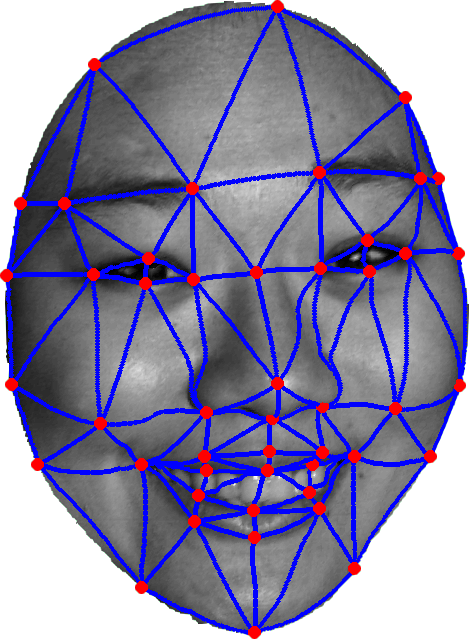} &
\includegraphics[height=2cm]{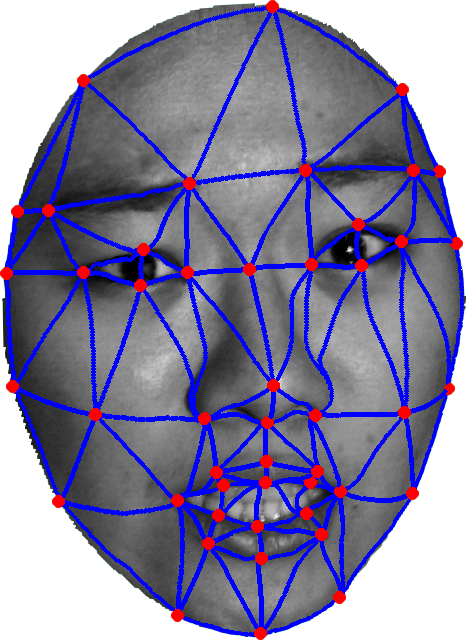} &
\includegraphics[height=2cm]{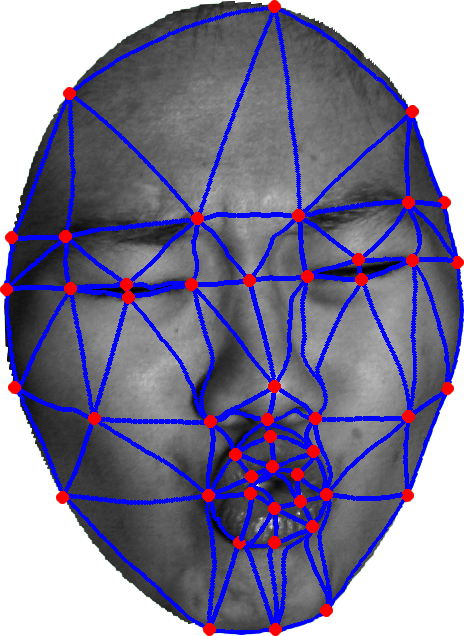} \\
neutral & smile & sad & pout \\
\includegraphics[height=2cm]{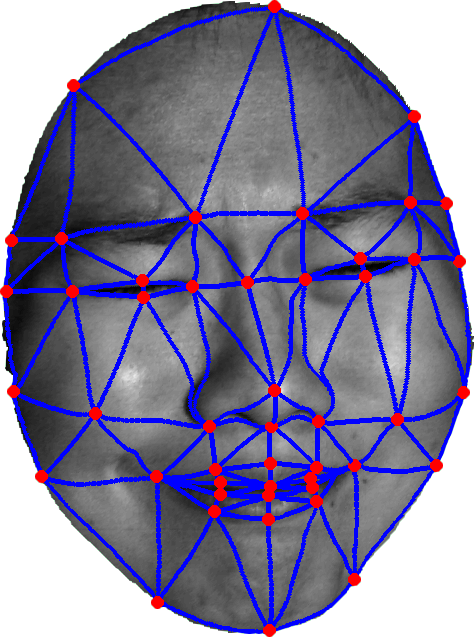} &
\includegraphics[height=2cm]{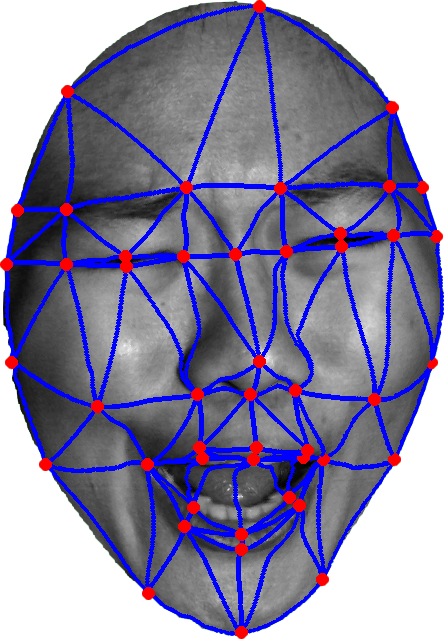} &
\includegraphics[height=2cm]{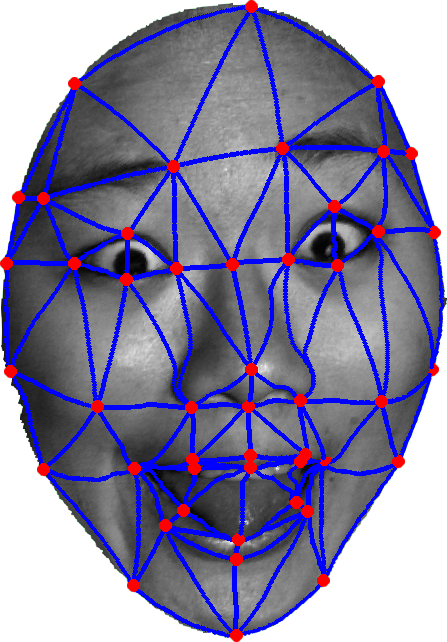} &
\includegraphics[height=2cm]{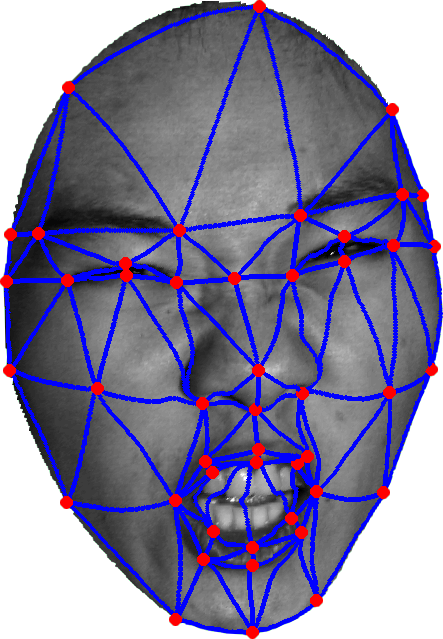} \\
bitter smile & pain & wry & ferocious
\end{tabular}
\caption{The geodesics on a human face with different facial expressions}
\label{fig:GeodesicRegistration}
\end{figure}

The effect of the traditional image morphing by using the direct interpolation is not satisfactory since there are shadows in the area of wrong correspondence. In 3D morphing, it could be even worse.
To improve this phenomenon, D. Smythe proposed the mesh warping technique \cite{W1998}, which partitions the 2D image into several pieces and interpolate them piece by piece.
Base on this idea, we compute a geodesic frame $\mathcal{M}$ on the surfaces in 3D. The frame $\mathcal{M}$ consists of the discrete surface geodesics that connect preselected feature points. The frame $\mathcal{M}$ can be refined to better capture characters of various facial expressions. The geodesic frames for the eight facial expressions are shown in Figure \ref{fig:GeodesicRegistration}.

\begin{figure}
\centering
\begin{tabular}{cccc}
\includegraphics[height=1.5cm]{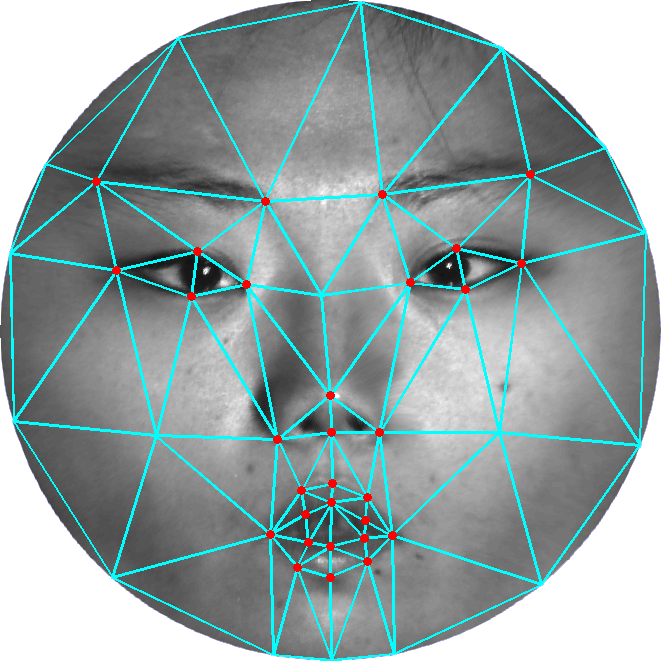} &
\includegraphics[height=1.5cm]{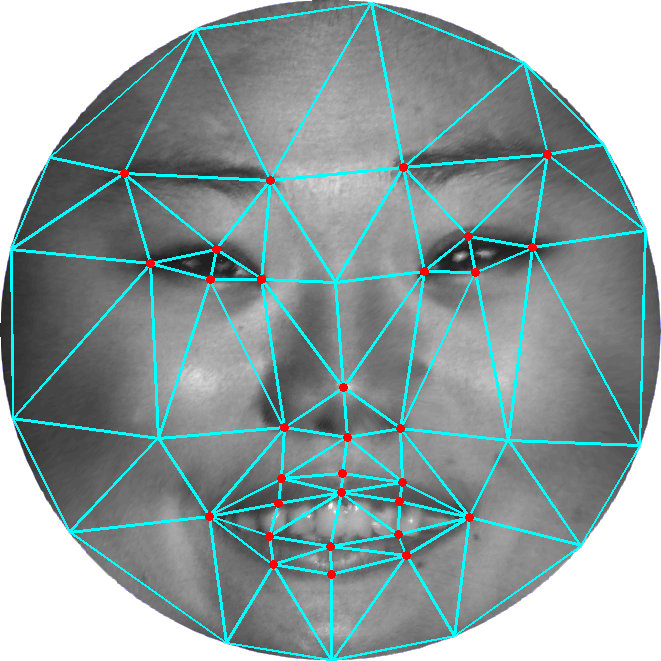} &
\includegraphics[height=1.5cm]{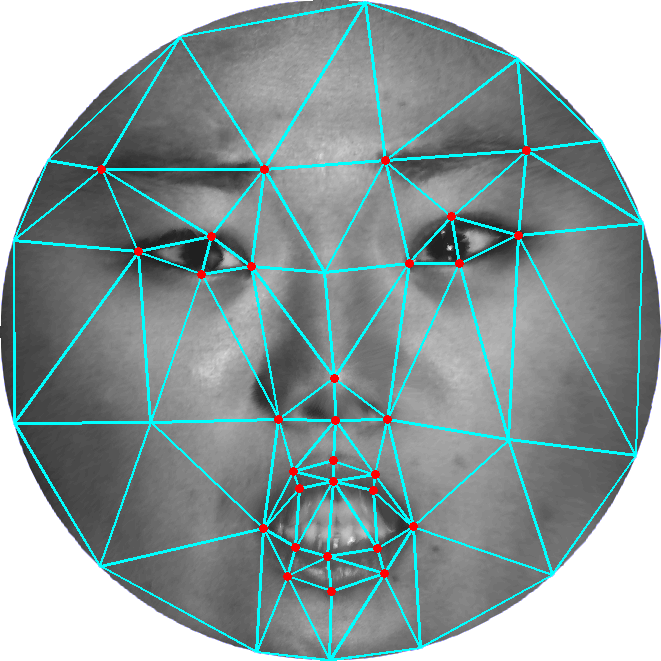} &
\includegraphics[height=1.5cm]{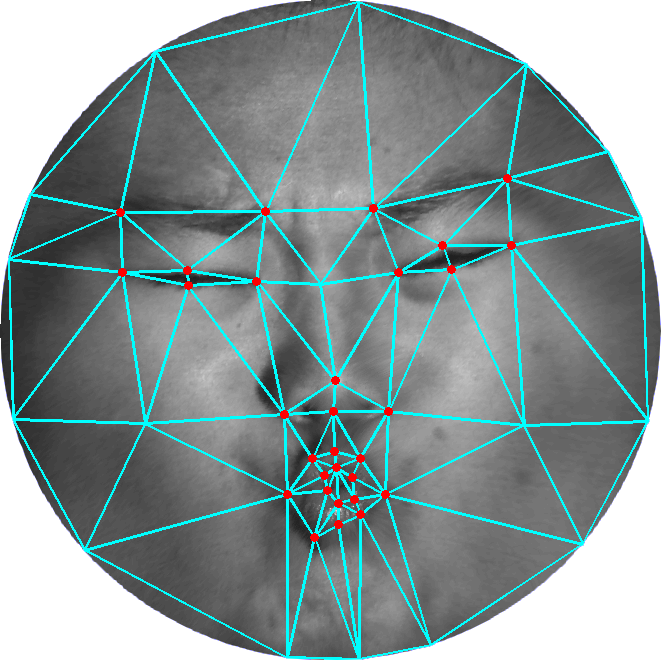} \\
neutral & smile & sad & pout \\
\includegraphics[height=1.5cm]{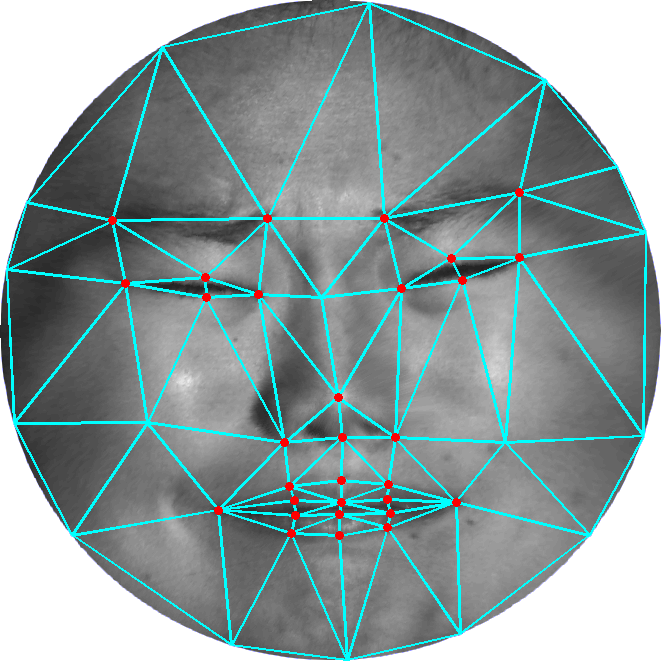} &
\includegraphics[height=1.5cm]{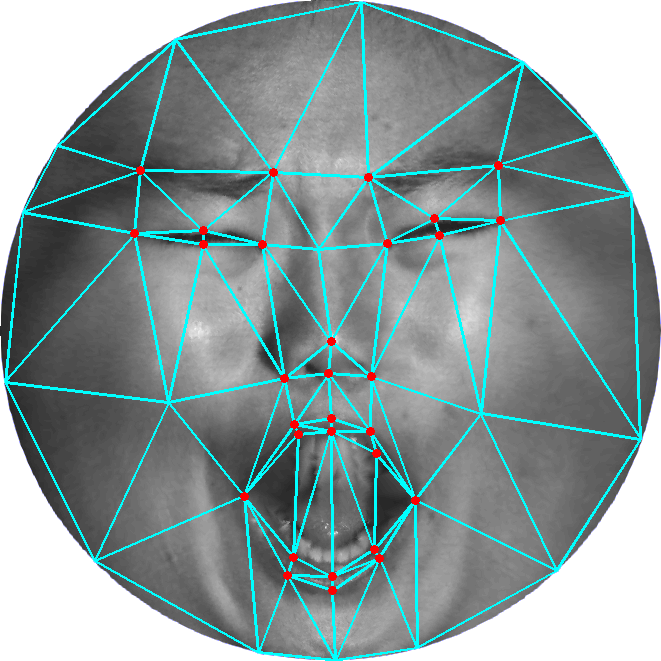} &
\includegraphics[height=1.5cm]{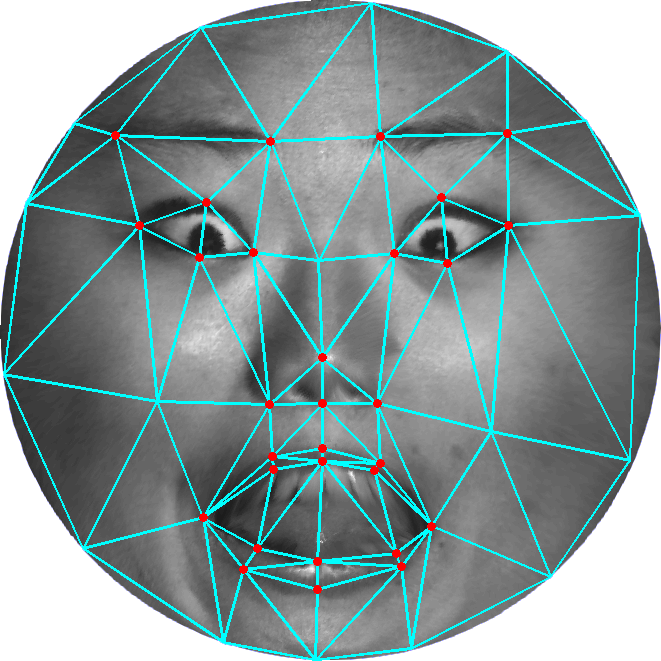} &
\includegraphics[height=1.5cm]{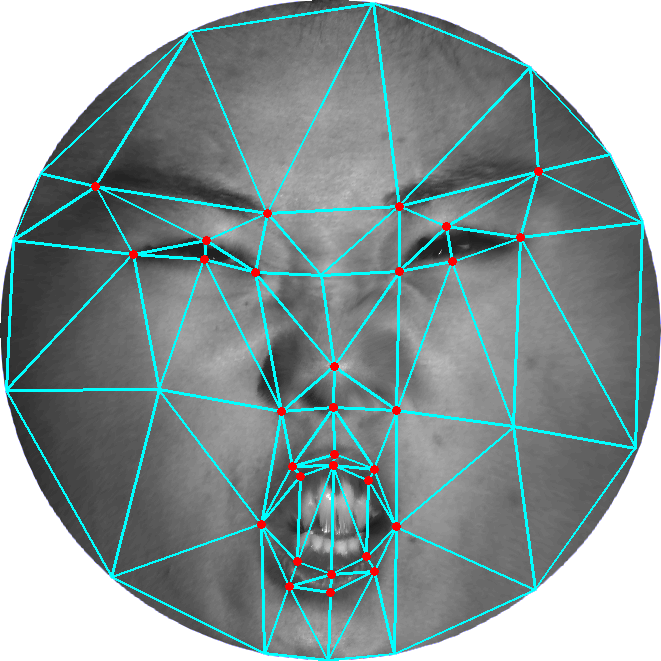} \\
bitter smile & pain & wry & ferocious
   \end{tabular}
\caption{The partition mesh of the unit disk for different facial expressions}
\label{fig:ConformalCoarseMesh}
 \end{figure}

The method of computing discrete surface geodesics on triangular meshes was proposed by D. Mart\'inez et al. \cite{2005MVC} which computes the initial path by using Sethian's fast marching method and correct the path with the path correcting iterations.
For a mesh with a large number of vertices, the fast marching method could be very time consuming. To improve the efficiency, we calculate the initial path by
\begin{itemize}
\item[(i)] construct the frame on the unit disk, where feature points are connected by straight line segments,
\item[(ii)] the initial paths are obtained by taking the inverse conformal map $\varphi^{-1}$ of these line segments.
\item[(iii)] Apply Mart\'inez's algorithm to obtain the geodesic frame.
\end{itemize}
The resulted geodesic frame is called the single mesh.
In our calculation, the initial paths in the frame mostly converge to the geodesics within 5 steps in the path correcting iterations.
The geodesic frame is remapped to partition the conformal parametric domain as shown in Figure \ref{fig:ConformalCoarseMesh}.

%%%%%%%%%%%%%%%%%%%%%%%%%%%%%%%%%%%%%%%%%%%
%%%%%%%%%%%%%%%%%%%%%%%%%%%%%%%%%%%%%%%%%%%

\section{Surface Registration via the Geodesic Frame of the Surface}
To build an one-to-one surface registration, we rely on the aforementioned partition mesh $\varphi(\mathcal{M})$. Let $\varphi_a(\mathcal{M}_a)$ be the partition mesh on $\varphi_a(S_a)$. The partition mesh on $\varphi_b(S_b)$ is determined by the $\overline{f}\circ\varphi_a(\mathcal{M}_a)$ where $\overline{f}$ is the matching function discussed in section \ref{Sec:SMMGE}.
To introduce our registration map, first let us introduce some notations.
The partition mesh of the unit disk is denoted by $\varphi(\mathcal{M})(\mathcal{V},\mathcal{F})$, here
\begin{align*}
\mathcal{V} &= \left\{ V_i \left| V_i = \left(V_i^{(1)}, V_i^{(2)}\right) \in \mathbb{D} \right. \right\}_{i=1}^{\#(\mathcal V)},
\end{align*}
and
\begin{align*} 
\mathcal F = \left\{ F_i \left| F_i =\left(F_i^{(1)}, F_i^{(2)}, F_i^{(3)}\right) \right.\right\}_{i=1}^{\#(\mathcal F)},
\end{align*}
are the set of vertices and the set of triangles in $\varphi(\mathcal{M})$
where $V_i^{(j)}, \,j=1, 2,$ is the coordinate of the $i$-th vertex and $F_i^{(j)}, \,j=1, 2, 3,$ is the indices of the vertices of the $i$-th triangle,
$\#(\mathcal V)$ and $\#(\mathcal F)$ denote the number of vertices in $\mathcal{V}$ and the number of triangles in $\mathcal{F}$.
Obviously, the closed region determined by $F_i$ can be easily represented by
\begin{align*}
\alpha_1(v) V_{F_i^{(1)}} + \alpha_2(v) V_{F_i^{(2)}} + \alpha_3(v) V_{F_i^{(3)}}, \\
\sum_{i=1}^3 \alpha_i(v) = 1,\,\, \alpha_i(v) \geq0, \,\, i=1,2,3,
\end{align*}
where $\left(\alpha_1(v), \alpha_2(v), \alpha_3(v)\right)$ is the barycentric coordinate of $v$ with respect to the vertices $V_{F_i^{(1)}}$, $V_{F_i^{(2)}}$ and $V_{F_i^{(3)}}$.  
A simple piecewise linear registration map  $\mathscr{R}_{\Phi}: S_a \to S_b$ can be easily constructed by the affine mapping
\begin{align*}
\Phi(v) = \alpha_1(v) \overline{f}(V_{F_i^{(1)}}) + \alpha_2(v) \overline{f}(V_{F_i^{(2)}}) + \alpha_3(v) \overline{f}(V_{F_i^{(3)}}),
\end{align*}
if $v\in F_i$, between each pair of triangles $F_i \in \varphi_a(\mathcal{M}_a)$ and $\overline{f}(F_i) \in \overline{f}\circ \varphi_a(\mathcal{M}_a)$,
where
\begin{align*}
\alpha_1(v) &= \frac{1}{2}\left\| \overrightarrow{v V_{F_i^{(2)}} } \times \overrightarrow{v V_{F_i^{(3)}} } \right\|_2,\\
\alpha_2(v) &= \frac{1}{2}\left\| \overrightarrow{v V_{F_i^{(3)}} } \times \overrightarrow{v V_{F_i^{(1)}} } \right\|_2,\\
\alpha_3(v) &= \frac{1}{2}\left\| \overrightarrow{v V_{F_i^{(1)}} } \times \overrightarrow{v V_{F_i^{(2)}} } \right\|_2.
\end{align*}

Recall $(H, \lambda)$ is a unique representation of $S.$
So, the surface registration can be realized through
$$\mathscr{R}_{\Phi}\left[ (H, \lambda)_a(v) \right] = (H, \lambda)_b(\Phi(v)).$$
Similarly, the texture images $\mathcal{T}_a$ and $\mathcal{T}_b$ of the surface $S_a$ and $S_b$, respectively, can be registrated by $$\mathscr{R}_{\Phi}\left[ \mathcal{T}_a(v) \right] = \mathcal{T}_b(\Phi(v)).$$

\begin{figure}[h!]
\centering
\begin{overpic}[height=2.5cm]{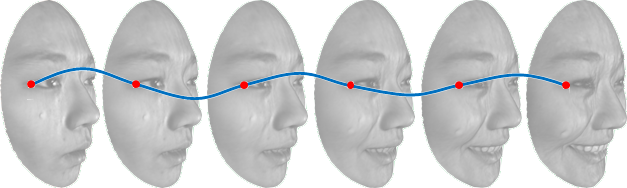}
\put(2.5,11){$(H, \lambda)_0$}
\put(19.5,11){$(H, \lambda)_1$}
\put(36.5,11){$(H, \lambda)_2$}
\put(54,11){$(H, \lambda)_3$}
\put(71,11){$(H, \lambda)_4$}
\put(88,11){$(H, \lambda)_5$}
\put(6,31){$S_0$}
\put(23,31){$S_1$}
\put(40,31){$S_2$}
\put(57,31){$S_3$}
\put(74,31){$S_4$}
\put(92,31){$S_5$}
\put(13,35){$\mathscr{R}_{\Phi_1}$}
\put(31,35){$\mathscr{R}_{\Phi_2}$}
\put(48,35){$\mathscr{R}_{\Phi_3}$}
\put(66,35){$\mathscr{R}_{\Phi_4}$}
\put(83,35){$\mathscr{R}_{\Phi_5}$}
\put(11,30){\includegraphics[width=1cm,height=0.3cm]{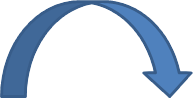}}
\put(28,30){\includegraphics[width=1cm,height=0.3cm]{images/arrow_map.png}}
\put(45,30){\includegraphics[width=1cm,height=0.3cm]{images/arrow_map.png}}
\put(63,30){\includegraphics[width=1cm,height=0.3cm]{images/arrow_map.png}}
\put(80,30){\includegraphics[width=1cm,height=0.3cm]{images/arrow_map.png}}
\end{overpic}
\caption{The cubic spline homotopy of the mean curvature and conformal factor of a vertex}
\label{fig:homotopy}
\end{figure}

In the following, we introduce how we utilize the above surface registration method to generate the morphing sequence through the cubic spline homotopy of the mean curvatures and the conformal factors.
Suppose 3D images of facial expressions $S_0, S_1 \ldots, S_{N},$ are captured a time $t_0,t_1,\ldots, t_N$.
Using the above surface registration method, the registration maps $\mathscr{R}_{\Phi_i}:S_{i-1}\to S_i,$ $i=1, 2, \ldots, N,$ can be easily computed. Using these registration maps, a morphing path $\textsl{P}(v,t)$, $t\in [t_0,t_N]$ and $v \in S_0$, can be created, here $\textsl{P}(v,t)$ denotes the location where a point $v \in S_0$ is morphed at time $t$. Since $(H, \lambda)$ is a unique representation of a surface, the morphing path can also be uniquely determined by the evolution of the conformal factor and the mean curvature. Here, we employee a piecewise cubic spline homotopy to interpolate
\begin{align*}
\left\{(H, \lambda)_0(v), (H, \lambda)_1(\Phi_1(v)), (H, \lambda)_2(\Phi_2\circ\Phi_1(v)),\right. \\
\left. \ldots, (H, \lambda)_N(\Phi_N\circ\Phi_{N-1}\circ\cdots\circ\Phi_1(v))  \right\}.
\end{align*}
A sketch in Figure \ref{fig:homotopy} illustrates this idea. Let $\mathscr{S}[x_1,\ldots,x_n](t)$ denote the piecewise cubic spline function with given data $x_1,\ldots,x_n$. The conformal factor $\lambda$ and the mean curvature $H$ at $\textsl{P}(v,t)$ can now be evaluated by
\begin{align*}
(H, \lambda)(\textsl{P}(v,t)) &= \mathscr{S}\left[(H, \lambda)_0(v), (H, \lambda)_1(\Phi_1(v)), \right.\\
&(H, \lambda)_2(\Phi_2\circ\Phi_1(v)), \ldots,\\
& \left. (H, \lambda)_N(\Phi_N\circ\Phi_{N-1}\circ\cdots\circ\Phi_1(v))  \right](t),
\end{align*}
Similarly, suppose the texture images $\mathcal{T}_{i}$ of the surface $S_{i}$, $i=1, \ldots, n,$ are given. The texture image associated with the surface along the morphing path $\textsl{P}(v,t)$ can be computed by the cubic spline homotopy
\begin{align*}
\mathcal{T}(\textsl{P}(v,t)) = \mathscr{S}&\left[\mathcal{T}_0(v), \mathcal{T}_1(\Phi_1(v)), \mathcal{T}_2(\Phi_2\circ\Phi_1(v)), \right.\\
&\left. \ldots, \mathcal{T}_N(\Phi_N\circ\Phi_{N-1}\circ\cdots\circ\Phi_1(v))  \right](t).
\end{align*}
Finally, by reconstructing the 3D surfaces $S_t, t\in [t_0, t_N],$ from their $(H, \lambda)$ representation and applying the computed texture image $\mathcal{T}_{t}$  to $S_t$, the morphing sequence between $S_0$ and $S_N$ can be obtained for any period from $t_0$ to $t_N$. In the following section, we shall introduce the surface reconstrution algorithm in detail.\\[0.5cm]

%%%%%%%%%%%%%%%%%%%%%%%%%%%%%%%%%%%%%%%%%%
%%%%%%%%%%%%%%%%%%%%%%%%%%%%%%%%%%%%%%%%%%

\section{Laplace-Beltrami Surface Reconstruction} \label{Sec:SR}
In the previous section, we have obtained the mean curvatures and the conformal factors $(H, \lambda)$ at each time period $t_i$, $i=1, 2, \ldots, n$. 
Gu and Yau proposed that any surface in 3D Euclidean space can be determined by its conformal factor and mean curvature uniquely up to rigid motions \cite{2004GWY}. 
We could generate the morphing sequence of surfaces $\{S_{t_i} = S(H, \lambda)_{t_i} \}_{i=1}^n$ by reconstructing the unique surface $S_{t_i}$ via $(H, \lambda)_{t_i}$, $i=1, 2, \ldots, n$.

In the following, we introduce how we utilize the mean curvatures and the conformal factors to reconstruct surfaces by solving the Laplace-Beltrami equations. 
Applying Gu and Yau's method, for the given mean curvature and the conformal factor $(H, \lambda)$, we reconstruct the unique 3D surface $S$ by solving the Laplace-Beltrami equations
\begin{equation*}
\left\{
\begin{aligned}
\Laplace_s S(u,v)&= 2H(u,v) \mathbf{n}(u,v)  \\
\frac{\partial S}{\partial u}(u,v) \times \frac{\partial S}{\partial v}(u,v) &= \lambda^2(u,v) \mathbf{n}(u,v)
S|_{\partial D} &=& \partial S
\end{aligned}
\right.,
\end{equation*}
where
$$\Laplace_s = \frac{1}{\lambda^2(u,v)}\left( \frac{\partial^2}{\partial u^2} + \frac{\partial^2}{\partial v^2} \right)$$
and $\mathbf{n}(u,v)$ is the normal of the surface $S$. 
The detail algorithm for solving the Laplace-Beltrami equations can be seen in Algorithm \ref{Alg:SurfaceReconstruction}.

\begin{algorithm}
\caption{Laplace-Beltrami Surface Reconstruction}
\label{Alg:SurfaceReconstruction}
\begin{algorithmic}[1]
\REQUIRE the mean curvature and the conformal factor $(H, \lambda)$ of the surface $S$ and the boundary $\partial S$ od the surface $S$.
\ENSURE the surface $S$.
\STATE Set the initial surface normal $\mathbf{n}^{(0)}.$
\REPEAT
\STATE Solve the boundary value problem $$\displaystyle\Laplace_s S^{(j+1)} = 2H \lambda^2 \mathbf{n}^{(j)},$$ where boundary $\partial S$ is known.
\STATE Update the surface normal $$\displaystyle\mathbf{n}^{(j+1)} = \frac{S_u^{(j+1)} \times S_v^{(j+1)}}{\lambda^2}.$$
\UNTIL{convergence}
\end{algorithmic}
\end{algorithm}

We reconstruct 7 frames from the video at time $t=0.0, 0.3, 0.6, 1.0, 1.3, 1.6, 2.0,$ respectively. The surface captured from the video at time $t$ is denoted by $S_t$ and the reconstructed surface $S(H, \lambda)_t$ is denoted by $\widetilde{S}_t$. 
In order to verify that the reconstructed surface is approximately the real surface, we compute the difference between the real surface and the reconstructed surface in $L^2$ sense
\[
\|S_t - \widetilde{S}_t \|_2 = \left( \int_{\mathbb{D}} \left| S_t(x) - \widetilde{S}_t(x) \right|^2 \diff x \right)^{\frac{1}{2}},
\]
and in $L^\infty$ sense 
\[
\|S_t - \widetilde{S}_t \|_\infty =  \max_{x\in\mathbb{D}}\left| S_t(x) - \widetilde{S}_t(x) \right|, 
\]
respectively. Table \ref{Tab:ReconstructionError} shows the reconstruction error in both $L_2$ norm and $L_\infty$ norm. 

\begin{table}
\centering
\begin{tabular}{c|c|c}
\hline
  t   & $\|S_t - \widetilde{S}_t \|_2$ & $\|S_t - \widetilde{S}_t \|_\infty$  \\
\hline\hline
$0.0$ & $3.8900\times 10^{-3}$ & $6.5598\times 10^{-4}$  \\
\hline
$0.3$ & $2.5601\times 10^{-3}$ & $3.8562\times 10^{-4}$  \\
\hline
$0.6$ & $4.6325\times 10^{-3}$ & $5.1009\times 10^{-4}$  \\
\hline
$1.0$ & $2.9394\times 10^{-3}$ & $3.9610\times 10^{-4}$  \\
\hline
$1.3$ & $4.4131\times 10^{-3}$ & $6.5401\times 10^{-4}$  \\
\hline
$1.6$ & $2.6350\times 10^{-3}$ & $6.6641\times 10^{-4}$  \\
\hline
$2.0$ & $4.6418\times 10^{-3}$ & $4.4197\times 10^{-4}$  \\
\hline
\end{tabular}
\caption{Reconstruction error between the real surface $S_t$ and the reconstructed surface $\widetilde{S}_t$}
\label{Tab:ReconstructionError}
\end{table}

%%%%%%%%%%%%%%%%%%%%%%%%%%%%%%%%%%%%%%%%%%%
%%%%%%%%%%%%%%%%%%%%%%%%%%%%%%%%%%%%%%%%%%%

\section{Numerical Results of Surface Morphing in 3D} \label{Sec:SM3D}

In the following, we show some surface morphing results via merely two images by using the surface morphing technique which we have mentioned above. 
In each figure, the left most image is the role of initial surface $S_a$ while the right most image is the role of terminal surface $S_b,$ and the images in the middle are the morphing sequence between $S_a$ and $S_b$. 

\begin{figure}
\includegraphics[height=1.8cm]{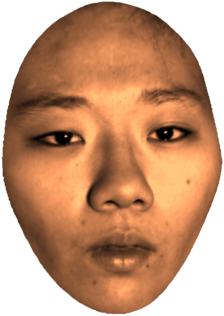}
\includegraphics[height=1.8cm]{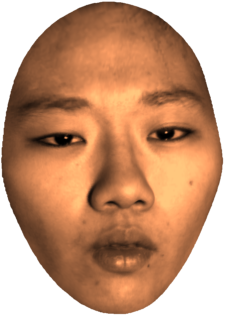}
\includegraphics[height=1.8cm]{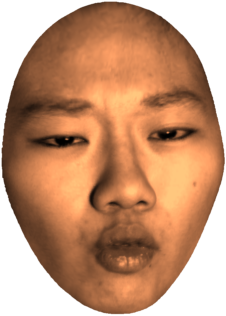}
\includegraphics[height=1.8cm]{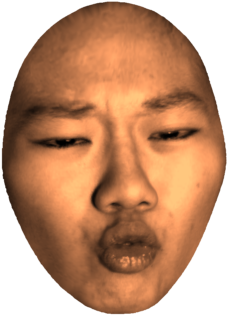}
\includegraphics[height=1.8cm]{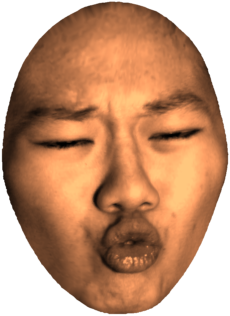}
\includegraphics[height=1.8cm]{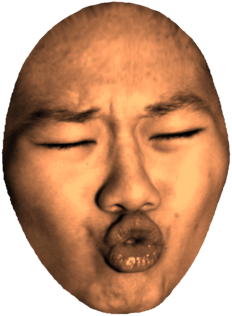}
\caption{The morphing sequence of eye blinking}
\label{EB}
\end{figure}

\begin{figure}
\includegraphics[height=1.8cm]{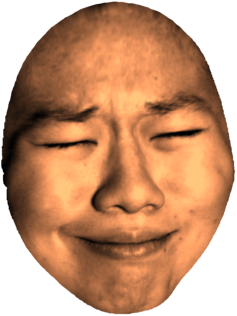}
\includegraphics[height=1.8cm]{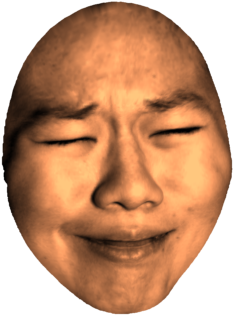}
\includegraphics[height=1.8cm]{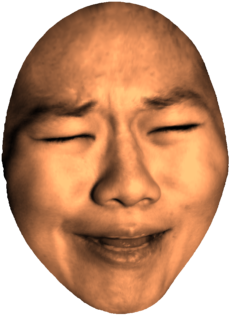}
\includegraphics[height=1.8cm]{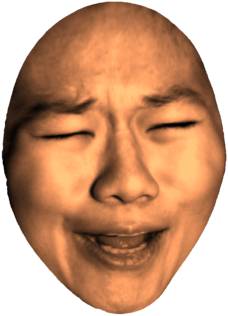}
\includegraphics[height=1.8cm]{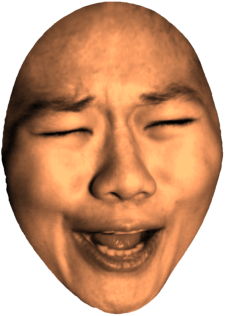}
\includegraphics[height=1.8cm]{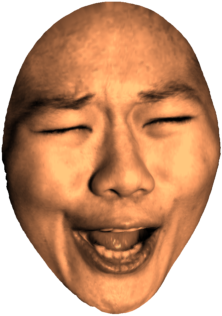}
\caption{The morphing sequence of mouth opening}
\label{MO}
\end{figure}

\begin{figure}
\centering
\includegraphics[height=1.8cm]{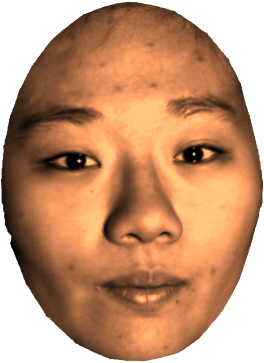}
\includegraphics[height=1.8cm]{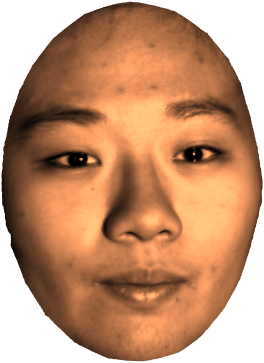}
\includegraphics[height=1.8cm]{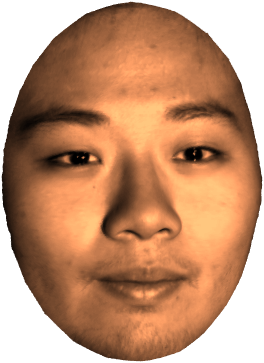}
\includegraphics[height=1.8cm]{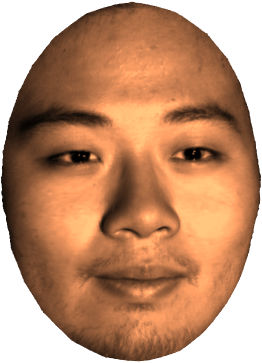}
\includegraphics[height=1.8cm]{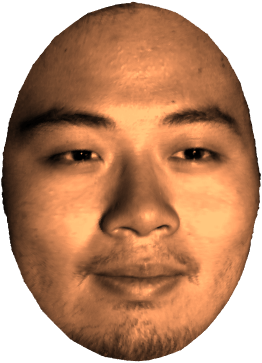}
\includegraphics[height=1.8cm]{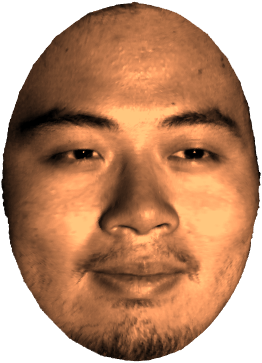}
\caption{The morphing sequence from a girl's face into a boy's face}
\label{fig:Morphingyylin2cyho}
\end{figure}

\begin{figure}
\centering
\includegraphics[height=1.6cm]{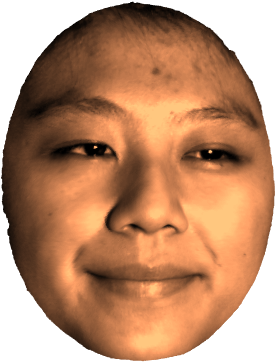}
\includegraphics[height=1.6cm]{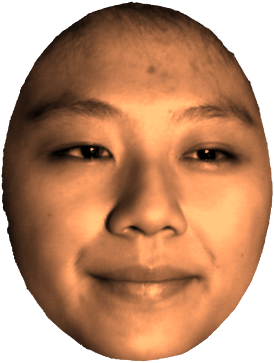}
\includegraphics[height=1.6cm]{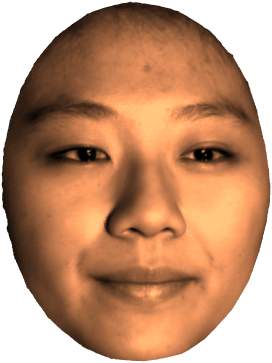}
\includegraphics[height=1.6cm]{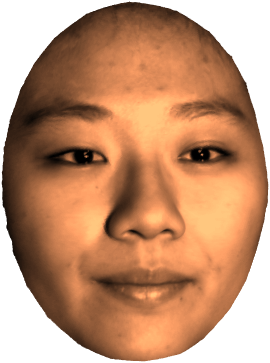}
\includegraphics[height=1.6cm]{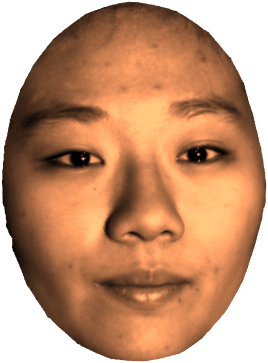}
\includegraphics[height=1.6cm]{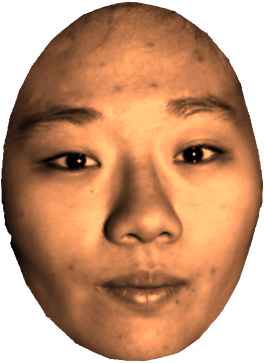}
\caption{The morphing sequence from a girl's face into another}
\label{fig:Morphingmshuang2yylin}
\end{figure}

\begin{figure}
\centering
\includegraphics[height=1.8cm]{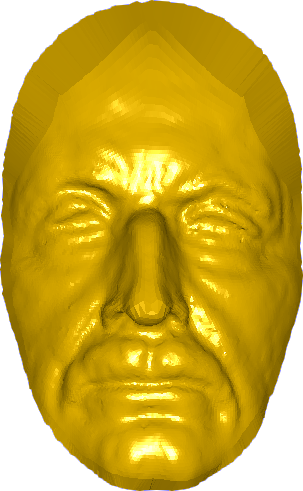}
\includegraphics[height=1.8cm]{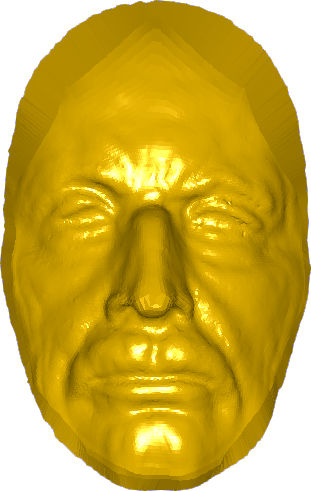}
\includegraphics[height=1.8cm]{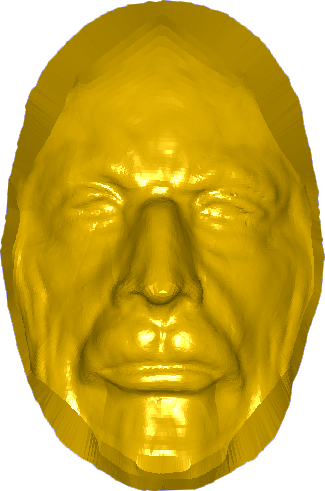}
\includegraphics[height=1.8cm]{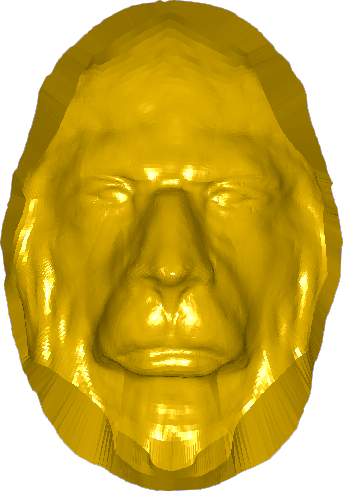}
\includegraphics[height=1.8cm]{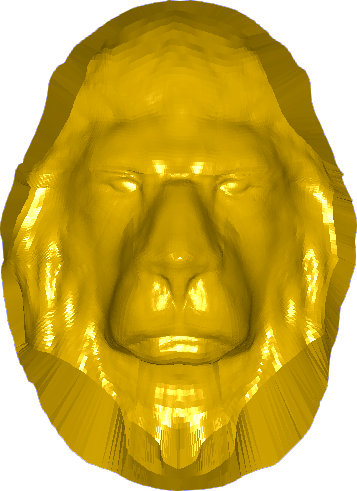}
\includegraphics[height=1.8cm]{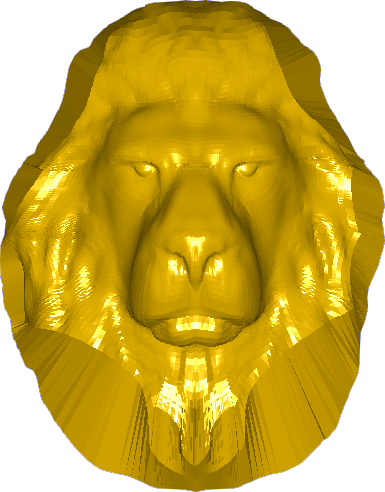}
\caption{The morphing sequence from Max Planck's face into a lion's head}
\label{fig:MorphingMax2Lion}
\end{figure}

Figure \ref{EB} and Figure \ref{MO} show the morphing sequence between two different facial expression, respectively. 
%Figure \ref{fig:MorphingEyes} shows the morphing sequence between  a face with closed eyes and a face with opened eyes. 
Figure \ref{fig:Morphingyylin2cyho} and Figure \ref{fig:Morphingmshuang2yylin} show the morphing sequence between faces of two different people, respectively. 
Figure \ref{fig:MorphingMax2Lion} shows the morphing sequence between a human face and a loin's head. 

It is interesting that how much the global matching function improves the morphing sequence. The affine mapping constructed by using the optimal M\"obius transformation is denoted by $\Phi_i^{M}$ and the affine mapping constructed by using the optimal M\"obius transformation with the global matching function is denoted by $\Phi_i^{G}$, $i=1,2$. 
In order to measure the rate of improvement, we reconstruct the surfaces $$\widehat{S}_t^{M} \equiv S(H, \lambda)_t^{M}$$ and 
$$\widehat{S}_t^{G} \equiv S(H, \lambda)_t^{G}$$
where 
\begin{align*}
(H, \lambda)_t^{M} = \mathscr{S}& \left[(H, \lambda)_0(v), (H, \lambda)_1(\Phi_1^{M}(v)),\right.\\
& \left. (H, \lambda)_2(\Phi_2^{M}\circ\Phi_1^{M}(v))\right](t),
\end{align*}
and 
\begin{align*}
(H, \lambda)_t^{G} = \mathscr{S}& \left[(H, \lambda)_0(v), (H, \lambda)_1(\Phi_1^{G}(v)),\right.\\ 
&\left. (H, \lambda)_2(\Phi_2^{G}\circ\Phi_1^{G}(v))\right](t),
\end{align*}
$t\in[0,2].$ Then, we compute the surface difference 
\[
\|\widehat{S}_t^M - S_t\|_2 = \left( \int_{\mathbb{D}} \left| \widehat{S}_t^M(x) - S_t(x) \right|^2 \diff x \right)^{\frac{1}{2}} 
\]
and 
\[
\|\widehat{S}_t^G - S_t\|_2 = \left( \int_{\mathbb{D}} \left| \widehat{S}_t^G(x) - S_t(x) \right|^2 \diff x \right)^{\frac{1}{2}}. 
\]
The rate of improvement is defined by 
\[
\frac{\|\widehat{S}_t^M - S_t\|_2-\|\widehat{S}_t^G - S_t\|_2}{\|\widehat{S}_t^M - S_t\|_2}.
\]
Table \ref{Tab:MorphingComparison} indicates that the global matching function improves approximately $50\%$ of the surface difference. 
\vspace{0.4cm}
\begin{figure}
\centering
\begin{overpic}[height=2cm]{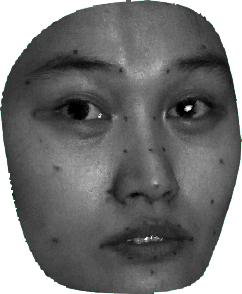}
\put(10,105){$S_0$}
\end{overpic}
\begin{overpic}[height=2cm]{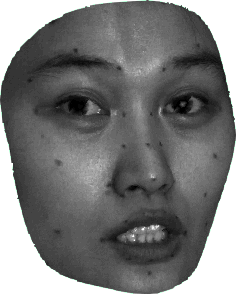}
\put(10,105){$S_{0.6}$}
\end{overpic}
\begin{overpic}[height=2cm]{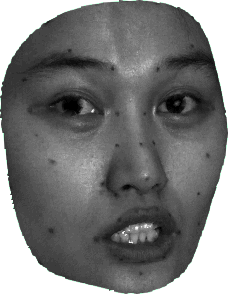}
\put(10,105){$S_1$}
\end{overpic}
\begin{overpic}[height=2cm]{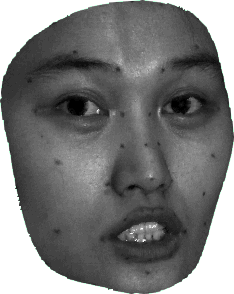}
\put(10,105){$S_{1.3}$}
\end{overpic}
\begin{overpic}[height=2cm]{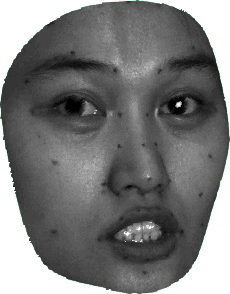}
\put(10,105){$S_2$}
\end{overpic}
\end{figure}
$~$\\[-1.2cm]
\begin{table}
\centering
\begin{tabular}{c|c|c|c}
\hline
 $t$   & $\|\widehat{S}_t^M - S_t\|_2$ & $\|\widehat{S}_t^G - S_t\|_2$ & Improvement Rate \\
\hline\hline
0.3 & $1.4980 \times 10^{-2}$ &  $8.6761\times 10^{-3}$ & 42.08\% \\
\hline
0.4 & $2.0554 \times 10^{-2}$ &  $1.1229\times 10^{-2}$ & 45.37\% \\
\hline
0.5 & $2.3328 \times 10^{-2}$ &  $1.0679\times 10^{-2}$ & 54.22\% \\
\hline
1.3 & $4.1535 \times 10^{-2}$ &  $1.8688\times 10^{-2}$ & 55.01\% \\
\hline
1.4 & $3.9139 \times 10^{-2}$ &  $1.9036\times 10^{-2}$ & 51.36\% \\
\hline
\end{tabular}
\caption{Comparison between the optimal M\"obius transformation and the optimal M\"obius transformation with the global matching function}
\label{Tab:MorphingComparison}
\end{table}

Moreover, extrapolation can also be achieved by using this homotopy technique. The real surface at time $t$ is denoted by $S_t$ and the reconstructed surface at time $t$ is denoted by $\widehat{S}_t \equiv S(H, \lambda)_t$ where
\begin{align*}
(H, \lambda)_t = \mathscr{S} &\left[(H, \lambda)_{0.6}(v), (H, \lambda)_1(\Phi^G_1(v)),\right.\\
& \left. (H, \lambda)_{1.3}(\Phi^G_2\circ\Phi^G_1(v))\right](t). 
\end{align*}
Table \ref{Tab:ExtrapolationError} shows the surface difference between the real surface $S_t$ and the reconstructed surface $\widehat{S}_t = S(H, \lambda)_t$ in $L^2$ sense
\[
\|S_t - \widehat{S}_t \|_2 = \left( \int_{\mathbb{D}} \left| S_t(x) - \widehat{S}_t(x) \right|^2 \diff x \right)^{\frac{1}{2}},
\]
and in $L^\infty$ sense 
\[
\|S_t - \widehat{S}_t \|_\infty =  \max_{x\in\mathbb{D}}\left| S_t(x) - \widehat{S}_t(x) \right|, 
\]
respectively, at time $t=1.4, 1.5, 1.6$ and $1.7.$ 
\begin{figure}
\centering
\begin{overpic}[height=2cm]{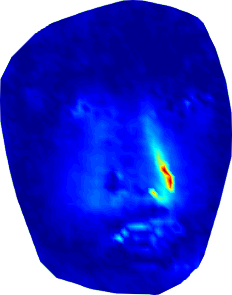}
\put(10,105){$S_{1.4}$}
\end{overpic}
\begin{overpic}[height=2cm]{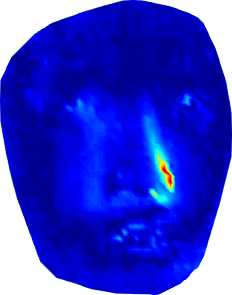}
\put(10,105){$S_{1.5}$}
\end{overpic}
\begin{overpic}[height=2cm]{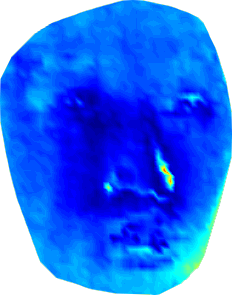}
\put(10,105){$S_{1.6}$}
\end{overpic}
\begin{overpic}[height=2cm]{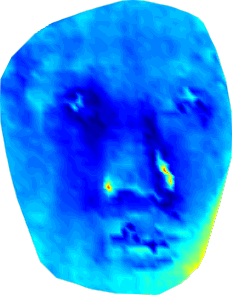}
\put(10,105){$S_{1.7}$}
\end{overpic}
\hspace{0.3cm}
\begin{overpic}[height=2cm]{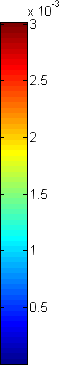}
\end{overpic}
\end{figure}
\begin{table}
\centering
\begin{tabular}{c|c|c}
\hline
  $t$   & $\|S_t - \widehat{S}_t \|_2$ & $\|S_t - \widehat{S}_t \|_\infty$  \\
\hline\hline
1.4 & $1.3639\times 10^{-2}$ & $2.9274\times 10^{-3}$  \\
\hline
1.5 & $1.2816\times 10^{-2}$ & $3.0208\times 10^{-3}$  \\
\hline
1.6 & $2.6365\times 10^{-2}$ & $2.4064\times 10^{-3}$  \\
\hline
1.7 & $3.2057\times 10^{-2}$ & $2.3978\times 10^{-3}$  \\
\hline
\end{tabular}
\caption{Extrapolation error between the real surface and the reconstructed surface}
\label{Tab:ExtrapolationError}
\end{table}

%%%%%%%%%%%%%%%%%%%%%%%%%%%%%%
\section{Conclusion}
In this paper, we proposed a 3D surface morphing method between different simply connected surfaces with single boundary in which smooth transient on both geometric characteristics and texture of the surfaces is considered. Similar to the traditional morphing approaches based on boundary representation, a wrap has to be created via feature correspondence and interpolation between shapes based on the wrap is employed to generate the morphing sequence. By taking advantage of the conformal parameterization and the unique surface representation of conformal factor and mean curvature, the wrap can be easily obtained by the composition of deformations from the M\"obius transformation and the thin-plate matching function. To mimic the non-isomorphic risk that usually occurs in matching largely deformed surfaces, a single mesh based on geodesic frame is employed. As a result, the correspondence, including geometric information and texture information, of the whole surface can be defined and interpolation among original surface and target surface can be computed by the usual cubic spline homotopy in a disk parametric domain. Finally, the morphing sequence can be generated from the surface reconstruction algorithm in section \ref{Sec:SR}. To make the proposed morphing approach more attractive in real applications, we improve the efficiency in computing the conformal parameterization and geodesic frames. We propose an non-linear iterative surface reconstruction algorithm (Algorithm \ref{Alg:SurfaceReconstruction}). The surface reconstruction algorithm can be accelerated by using the multigrid method on a uniform mesh by which multi-resolution surfaces can also be obtained. Several morphing effects among different 3D facial expressions are presented to demonstrate the feasibility of the proposed morphing method.

\ifCLASSOPTIONcaptionsoff
  \newpage
\fi

\end{document}